\documentstyle[epsfig]{mn}

\newcommand{ \ho }{$ H_{0} $ } 
\newcommand{ \hoc }{ $ H_{0} $}
\newcommand{ \mpc}{ h$^{-1}$Mpc}
\newcommand{ \mpcs}{ h$^{-1}$Mpc }
\newcommand{ \km }{kms$^{-1}$}
\newcommand{ \kmM }{ kms$^{-1}$Mpc$^{-1}$ }
\newcommand{ \omeg }{$ \Omega_{0} $ }
\newcommand{ \omegc }{$ \Omega_{0} $}
\newcommand{ \B }{$ \beta $ }
\newcommand{ \ci }{$ \chi^{2} $}

\begin{document}

\title[Peculiar velocities of nearby galaxies]{ Predicting the peculiar 
velocities of nearby PSC-z galaxies using the Least Action Principle  }

\author[J. Sharpe et al.]
{J. Sharpe,$^1$ M. Rowan-Robinson,$^1$ A. Canavezes,$^1$ 
W. Saunders,$^2$ \cr E. Branchini$^3$,
G. Efstathiou,$^4$ C. Frenk,$^3$ O. Keeble,$^1$  
R. G. McMahon,$^4$ \cr S. Maddox,$^4$  S. J. Oliver,$^1$ 
W. Sutherland,$^5$ H. Tadros$^{6,7}$ and S. D. M. White.$^8$ \\
$^1$Astrophysics Group, Blackett Laboratory, Imperial College of Science Technology and Medicine, \\ Prince Consort Road. London SW7 2BZ, UK \\
$^2$Institute for Astronomy, University of Edinburgh, Blackford Hill, 
Edinburgh EH9 3JS, UK\\
$^3$Department of Physics, University of Durham, South Road, 
Durham, DH1 3LE, UK\\
$^4$Institute of Astronomy, University of Cambridge, Madingley Road, 
Cambridge CB3 OHA, UK\\
$^5$Department of Physics, University of Oxford, Keeble Road, 
Oxford OX1 3RH, UK \\
$^6$Department of Physics, University of Sussex, Falmer, 
Brighton BN1 9QH, UK \\
$^7$ Present Address: Department of Physics, University of Oxford, 
Keeble Road, Oxford OX1 3RH, UK \\
$^8$Max-Planck-Institut f\"{u}r Astrophysik, 
Karl-Schwarzschild-Stra\ss{}e 1, 
85740 Garching bei M\"{u}nchen, Germany}

\date{received,  accepted} 

\maketitle

\begin{abstract}
We use the Least Action Principle to predict the 
peculiar velocities of PSC-z galaxies inside $ cz=2000 $\km. Linear theory 
is used to account for tidal effects to  $ cz=15000 $\km, and we iterate
galaxy positions to account for redshift distortions. As the     
Least Action Principle is valid beyond Linear theory, we can predict 
reliable peculiar 
velocities even for very nearby galaxies (ie $ cz \leq 500$\km).
These predicted peculiar velocities are then compared with the observed 
velocities of $12$ galaxies with Cepheid distances. The combination of the 
PSC-z galaxy survey (with its large sky coverage and uniform 
selection), with the accurate Cepheid distances, makes this comparison 
relatively free from systematic effects. We find that galaxies are 
good tracers of the mass, even at small ($ \leq 10$\mpc) scales; 
and under the assumption of no biasing, 
$ 0.25 \leq \beta \leq 0.75 $ (at $90\%$ confidence). We use the reliable
predicted  peculiar velocities to estimate the Hubble constant \ho 
from the local volume 
without ``stepping up'' the distance ladder, finding a confidence range 
of $ 65-75$\kmM   (at $90\%$ confidence).    
\end{abstract}

\begin{keywords}
galaxies:distances and redshifts - dark matter - distance scale
- large-scale structure of universe- methods:numerical
\end{keywords}

\section{Introduction}

The relationship between galaxies and matter is one of the most important 
questions in cosmology. On large scales, there is good evidence that
galaxies (and in particular IRAS galaxies) are good tracers of the mass.
For example: there is the good alignment of the IRAS and CMB dipoles
(Rowan-Robinson et al 1990, Strauss et al 1992), the agreement between 
predicted peculiar velocities  and Tully-Fisher velocities 
(Willick et al 1997a), and the consistency of POTENT density fields with
density fields from galaxy surveys (Sigad et al 1997). 
However, on smaller scales, the reliability of galaxies
as tracers of the mass is far from certain.

Investigations of peculiar velocities in the Universe have relied 
heavily on use of the linear term in gravitational instability theory
(eg, Rowan-Robinson et al 1990, Strauss et al 1992, Willick et al 1997a,
da Costa et al 1998). This has the advantage of being easy and computationally 
quick, but has the side effect of restricting such analysis to large scales
$ \sim 10 $\mpc, where linear theory is valid. In this paper we use
 the Least Action Principle, which is valid beyond linear theory. It is
therefore possible to predict peculiar velocities for nearby galaxies.
The Least Action Principle is similar to the more usual action principle, 
but with the conditions that the initial conjugate momenta are zero and 
the final positions fixed, instead of the more usual conditions
 of both the initial
and final positions being fixed. It was first used by 
Peebles (1989(P89), 1990(P90)),
to study the dynamics of the Local Group. Later papers Peebles (1994,1995),
presented different versions of the method, but still dealt mainly 
with the problem of the Local Group, with only a few external masses.
Shaya et al (1995(S95)), 
used the method to predict motions for galaxies inside
$ cz=3000$\km, and our methodology is very similar to theirs. 
Recently, Schmoldt \& Saha (1998), presented a version of the method directly
applicable to redshift space.

Previous work on the local velocity field at small scales has centred mainly on
trying to mimic the observed ``cold flow'' (low value for the rms of 
galaxy velocities) of local galaxies in N-body
simulations (Schlegel et al 1994, Governato et al 1997). In both papers
it was found to be virtually impossible for an $ \Omega_{0}=1 $ CDM model
to produce a flow as cold of that of the local galaxies.
In this paper we will compare the observed peculiar velocities of galaxies 
with cepheid distances, to their predicted peculiar velocities from the 
Least Action Principle. 
This more direct comparison will have the 
advantage of being less model dependent.
The emphasis of direct comparisons of predicted and observed velocity fields
has traditionally been on more distant galaxies. 
This is due to the wish to use 
linear theory and the need to have a sufficient number of galaxies to combat 
the large random errors associated with Tully-Fisher distance measurements.
Our analysis is similar to the one by S95, although we restrict our attention
to a local sample of galaxies ($7/12$ of our distances are less than $10$Mpc).
We also offer two major improvements.
Firstly the PSC-z redshift survey has the
advantages of greater uniformity and larger sky coverage, than the
optical catalogue of S95. Secondly the greater
availability of Cepheid distances, gives us a far superior 
(though smaller) set of distances. Both of these improvements 
should leave our estimate of \B and other results freer of systematic
errors.

It is difficult to measure \ho in the local volume, due to the effects
of peculiar velocities on the redshifts of galaxies. The traditional 
approach to this problem has been to extend the distance scale out to 
distances where the peculiar velocities will be negligible compared 
to the Hubble flow. However, the reliable peculiar velocities which we
predict, allow us the possibility of measuring \ho locally, without 
``stepping up'' the distance ladder. Such a measurement will obviously be a 
useful check on more traditional distance ladder determinations.

The layout of this paper is as follows. In the next Section we discuss the 
Least Action Principle. In Section \ref{mtd} we discuss our mass tracers
(mainly the PSC-z galaxy redshift survey) 
and the distance estimations we have used. 
In Section \ref{resvel} we discuss the results from the peculiar 
velocity comparison. Finally, in Section \ref{dis} 
we give a discussion of our results.

\section{The Least Action Principle}
\label{nam}

 The action S for a system 
of point masses in an expanding universe is (P89,P90,S95),     
\begin{equation}
S = \int_{0}^{t_{0}} dt L = \int_{0}^{t_{0}} dt \left[ \sum_{i} 
 \frac{m_{i} a^{2} \dot{ {\bf x}_{i}^{2} }}{2} - m_{i} 
\phi( {\bf x}_{i} )
 \right].   \label{act}
\end{equation}
Where $a$ is the expansion factor for the Universe ($a_{0}=1$), $ m_{i}$ 
is the mass of the $i$th particle, ${\bf x}_{i}$ its position and
$\phi( {\bf x}_{i})$ is the gravitational potential.

The action principle is derived by considering changes to the action under
infinitesimal changes to the orbits $ {\bf x}_{i}= {\bf x}_{i} + 
\delta {\bf x}_{i}(t) $, and then integrating equation {\ref{act} by parts,
\begin{eqnarray}
\delta S   = &  \int_{0}^{t_{0}} \delta t  \left[ \sum_{i} 
\delta {\bf x}_{i} \cdot  \left(  -  \frac{d}{dt} m_{i} a^{2} 
\frac{ \delta {\bf x}_{i}}{ \delta t} + m_{i} \frac{ \delta \phi 
( {\bf x}_{i} )}{\delta {\bf x}_{i}} \right) \right] \\ \nonumber
  &  +   \sum_{i} \left[ m_{i} a^{2} \delta {\bf x}_{i} \cdot 
\frac{ d {\bf x}_{i}}{dt} \right]_{0}^{t_{0}}. \label{dact}
\end{eqnarray} 
To get the equations of motion from the stationary points of the action,
we require the second term to vanish. 
Normally this is done by setting $ \delta {\bf x}_{i} =0 $ at the initial 
and final positions. 
However, it is also possible to apply the action principle with the
boundary conditions $ \delta {\bf x}_{i} = 0 $, at the final position and 
$ a^2 \delta {\bf x}_{i} / dt =0 $, at the initial position. This condition
will be satisfied by the growing mode in linear theory and is a natural
condition for the early universe.

We now have to find stationary points of the action. In practice, we will 
only look for minima. The global minimum will correspond to the solution 
closest to the Hubble flow, whist other extrema will apply to solutions where 
(for example) masses have crossed once and our now falling back
 into each other. When Peebles (P90) applied the method to the
Local Group, only one galaxy (N6822) required a non-minimum solution to be 
well modelled.  
To find the minima numerically we expand the galaxy orbits in terms
of a set of trial functions,
 \begin{equation}
x_{i}^{ \alpha} (t) = x_{i}^{ \alpha } (t_{0} ) + \sum_{n} C_{i,n}^{ \alpha} f_{
n} (t).
\end{equation}
Then to find the orbits we minimise the action with respect to the
coefficients $ C_{i,n}^{ \alpha} $. The choice of expansion functions 
$ f_{n} (t) $, is in theory arbitrary; as long as they satisfy the boundary
conditions. Though their growth rate should be consistent with linear theory 
at early epochs (ie $adx/dt \rightarrow a^{1/2}$ at $a \rightarrow 0$).
We follow previous papers (eg P89, P90, S95), 
in using the set,
\begin{equation}
f_{n} (t) = (1-a)^{n} a^{5-n} \left[ \frac{ 5!}{ n! (5-n)!} \right],  
\quad 1 \leq n \leq 5.
\end{equation}    
The equal spacing of the peaks of these functions allow for good convergence
and their relatively simple form makes them easy to use. 
The coefficients are found by ``walking down''; that is by following the 
gradient of the action towards the minimum. Normally we start with all 
$ C_{i,n}^{ \alpha}=0 $ and iterate according to 
the equation,
\begin{eqnarray}
 m_{i} \delta C_{i,n}^{ \alpha } & \propto & - \frac{ \delta s}
{ \delta C_{i,n}^{ \alpha} } \\
   & \propto & -m_{i} \int^{t_{0}}_{0} dt (   a^{2} 
\frac{ d x^{ \alpha}_{i}} {dt} \frac{d f_{n}}{d t} 
+ a f_{n} g^{\alpha}_{i}). \label{cit}
\end{eqnarray}
Where $g^{\alpha}_{i}$ is given by,
\begin{equation}
g^{\alpha}_{i} =  \frac{4}{3} \pi G \rho a {\bf x}^{\alpha}_{i}
+ \frac{G}{a^{2}} \Sigma_{j} m_{j} \frac{{\bf x}^{\alpha}_{j} -
 {\bf x}^{\alpha}_{i}}{ | {\bf x}_{j} -{\bf x}_{i} |^3 }. \label{gip}
\end{equation} 
Where $\rho$ is the mean density of the universe. 
This process continues for a sufficient number of iterations 
(see Section \ref{nitty}), that the $ C_{i,n}^{ \alpha} $ have converged
to near constant values. We will then know the orbit of every mass and thus 
its peculiar motion. We place all of the available mass in the point masses 
and unlike previous papers (P90, S95), do not consider the possibility of a 
smooth component to the mass (which was found to have little effect 
on the results).  

So far our discussion has been in terms of point masses. In reality, though,
we will have to apply the Least Action Principle to a set of galaxies. As
a result we will have to make the assumption that all of the mass is 
concentrated in these galaxies.
 An important consequence
of this assumption is that we require the size of galaxy halo
(or group halo, see Section \ref{group}) to be less than the average distance
between galaxies. There is some support for this approximation from studies
of the Local Group. Zaritsky et al (1989), apply the timing argument to
the Milky Way- Leo I system and the Milky Way- M31 system, and find 
roughly the same mass for the Local Group. As Leo I is at $200kpc$ and
M31 is at $700kpc$, this implies that the Milky Way does not have a large
extended halo. In addition, Peebles (1994,1995) finds that it is possible to 
construct satisfactory orbits for Local group satellites and nearby groups
with simple compact haloes.
 Another consequence of the assumption that all of the mass is 
concentrated in galaxies
is to place limits on the distribution of loose mass in 
the Universe (by loose mass we mean mass which is not in haloes).
 This is particularly important in CDM models where up to $50$ \%
of the mass can be loose in a $ \Omega_{0}=1$ universe and greater
proportions for lower density universes (Governato, private communication).

At the scale of the Local Group, 
Branchini \& Calberg (1994) and Dunn \& Laflamme (1995) have tested the
Least Action Principle in CDM universes.      
The results from these tests are not totally reliable, as the simulations 
used had either been rescaled for use on Local Group, or lacked resolution.  
Both papers found a tendency for the Least Action Principle to 
under-predict \omegc.
Branchini \& Calberg attributed these problems to extended haloes,
whereas  Dunn \& Laflamme thought the problem was more to do with 
loose mass. Our analysis is at larger scales, so it is not clear 
what to expect in CDM models.  

The condition that the mass is concentrated at the centre of the 
mass tracers has no chance of being satisfied at early epochs.
The best we can hope for is that the perturbations are roughly spherical
and centred on the mass tracers. We are, though, entirely concerned
with the velocities of galaxies and these will be generated mainly at 
late epochs. Thus the problems with the tracing of the density field at 
early epochs should not be too serious.

\section{Mass Tracers and Distances}
\label{mtd}

In this Section, we discuss the galaxy catalogues to which we apply the Least 
Action Principle. However, raw galaxy catalogues in the form of a list of 
galaxy positions, redshifts and flux, are not suitable to have the Least 
Action Principle directly applied to them. For example, the raw redshifts 
will have a contribution both from the Hubble flow and the peculiar velocity 
of a galaxy, so will not be the true measure of distance we require. 
Consequently, this Section will include not only a description of the galaxy 
catalogues we use, but also a description of the processes we apply to the 
galaxy catalogues, to make them suitable for calculations with the 
Least Action Principle.    

For peculiar velocity work, there will be certain desirable properties 
we will want in a galaxy catalogue. The most important of these is that the 
catalogue is uniformly selected across the sky, as preferential selection in 
any region, will result in false motions being calculated towards that 
region. Another important property is that we want the catalogue to cover as 
great a fraction of the sky as possible. All catalogues will have an 
``excluded region'' or ``mask'' around the galactic plane. 
In this region it is 
impossible to select galaxies reliably. In optically selected surveys 
this is because of absorption from dust. 
In infrared selected surveys due to cirrus emission.
It will clearly be desirable for this excluded region to be as small as 
possible. The last desirable property is simple density of sampling. 
The more galaxies we have in a given volume, the greater chance 
we will have of 
calculating accurate peculiar velocities; assuming our galaxies are 
reliable tracers of the mass. 

In Subsection \ref{subias}, we discuss how galaxy bias effects 
our analysis. We then describe the PSC-z survey, our main source of galaxies.
Subsection \ref{pscz} is a general introduction to the catalogue and is 
followed by Subsections on: the masked regions, the grouping of galaxies, 
the Virgo Cluster, tidal effects and redshift distortions. In Subsection 
\ref{locgal}, we discuss how we have treated local galaxies (those with 
$ cz \leq 500$\km). Finally, in Subsection \ref{dist}, we present the distance 
estimations we have used to derive the observed peculiar velocities.  

\subsection{ Galaxy Biasing in our Analysis}
\label{subias}

A preliminary issue to be discussed is the effect of bias in the 
galaxy distribution on our analysis.
We need a catalogue of mass tracers on which to apply the Least Action 
Principle, this will be a catalogue of galaxies and galaxy groups.
This means that our analysis will depend on the relationship between
the galaxies and the mass. In addition to our requirement that the mass
is concentrated in the galaxies (see Section \ref{nam}), we have
uncertainty from whether the mass associated with galaxies is dependant 
on their environment. 
This is the problem of the  bias of the galaxy distribution.
The standard approach to this problem is to use the linear bias factor $b$,
to quantify the uncertainty in this relationship
($b$ is given by $\delta_{g}=b \delta_{m}$, where $\delta_{m}$ and $\delta_{g}$ are the over- or under-densities in the galaxies and the mass respectively).
This has the advantage that in linear theory, the effect of $b$ 
is degenerate with that of \omeg and the 
results only depend on $\beta=\Omega^{0.6}_{0}/b$. 
However, the Least Action Principle does not rely on linear theory
and one of the main reasons for using it is to probe small scales at which 
linear theory is invalid. Thus we can no longer expect \omeg and $b$ to be
fully degenerate. 
We chose to weight our galaxies on the assumption that there is no bias 
(ie $b=1$). This is not as restrictive as it might appear. Although strictly
\omeg and $b$ are not degenerate, we expect that even at non-linear scales
a large degeneracy will persist. Support from this comes from 
(Dekel et al 1993), who investigated the peculiar velocity field using the
POTENT methodology. This methodology included non-linear effects and could in 
principle break the degeneracy between \omeg and $b$. They found, though, the 
individual constraints found on \omeg and $b$ were extremely weak and only the 
parameter \B was constrained to good accuracy.

The POTENT analysis is at fairly large scales, with the galaxy and velocity 
fields being smoothed on 1200 \km. Our analysis is at smaller scales and is 
thus more likely to see the degeneracy between \omeg and $b$ broken. 
Consequently, we have investigated the effect of bias in our calculations.
We can do this by modifying the expression used for the potential in equation
\ref{act}. Including bias, the potential can be expressed as 
\begin{equation}
\phi(x) = \left( \frac{ \rho_{mb}}{\rho_{b}} \right)   
\frac{-Ga^{2}}{b} \int d^{3}x' \frac{\rho (x')- \rho_{b}}{|x'-x|}. \label{bpot}
\end{equation}
Where $\rho(x')$ is the density of galaxies at x' and $\rho_{b}$ is the 
background density of galaxies. $\rho_{mb}$ is the background density of matter
and the factor $(\rho_{mb} / \rho_{g})$ just introduces a physical mass 
(as oppossed to galaxy number) into the equation. For b=1, this reduces to 
the standard form for the potential (Peebles 1980). Using equation \ref{bpot}
for the potential, equation \ref{gip} becomes,  
\begin{equation}
g^{\alpha}_{i} =  \frac{4}{3} \pi G \rho a {\bf x}^{\alpha}_{i}
+ \frac{G}{ba^{2}} \Sigma_{j} m_{j} \frac{{\bf x}^{\alpha}_{j} -
 {\bf x}^{\alpha}_{i}}{ | {\bf x}_{j} -{\bf x}_{i} |^3 }. \label{gipb}
\end{equation}
We can then use equation \ref{gipb} in equation \ref{cit} and find the minima 
of the action for a biased galaxy distribution 
and thus investigate the effect of 
bias. We should note, though, that our treatment is not entirely satisfactory. 
We have introduced bias in equation \ref{bpot} in a manner that is reliant on 
smooth and continuous density fields. However, the Least Action Principle 
applies to a set of point masses not continuous fields. We have thus 
introduced bias in an inconsistent manner. To have introduced bias 
consistently would of required us to weight galaxies according local density.
Such a process would be both complicated and noisy and thus we have chosen 
not to pursue it. We can, though, expect an investigation of bias using 
equations \ref{bpot} and \ref{gipb} to give at least an indication of
its effect on our results.

We thus did a series of runs with one of our input catalogues with different 
\omeg and $b$ but constant \B (the catalogue 
we used had a PSC-z local weighting and a cloned mask see Subsections 
\ref{pscz} - \ref{locgal}).
We then compared the magnitudes of the predicted velocities in the different 
calculations. For $\beta=0.5$ ($ \Omega_{0}=0.3$ $b=1$) 
then reducing $b$
to 0.5 ($\Omega_{0}=0.1$) resulted in the velocities decreasing on average 
to a factor 0.77 of the $b=1$ velocities. Whilst, increasing $b$ to 2 increased
the velocities to a factor $1.29$ of the $b=1$ velocities. Given these results,
it is clear that we still have a large degeneracy between \omeg and $b$
and the correct quantity for us to estimate is \B. Indeed, with the fairly 
weak limits we will eventually obtain on \B (see Subsection \ref{cl}) it would 
be futile to attempt to estimate \omeg and $b$ independently. Thus we will 
only estimate \B and not \omeg and $b$. We will, though, quote are initial 
results in terms of \omeg as we do not wish to hide the assumption that $b=1$,
used in obtaining these results.

An additional consideration comes from the possibilty of non-linear bias.
If non-linear terms are present in the biasing relation, then they are 
predicted to be more important at smaller scales 
(Mo \& White 1996) and so could
potentially be important for our work. However, most non-linear biasing 
models, tend to differ substantially from linear bias only
in clusters or voids (see the models of Mann, Peacock \& Heavens 1998). 
The local volume is neither of these
(Kraan-Korteweg \& Tammann 1979), so for our analysis most non-linear 
biasing schemes will be consistent
with linear bias and our weighting scheme. As a result our conclusions
on how galaxies trace the mass, will be for a general trend and not
a particular biasing scheme.

\subsection{ The PSC-z Survey}
\label{pscz}

The galaxy catalogue we use is the PSC-z survey (Saunders et al 1996). 
This is a redshift survey to the 
full depth ($S_{60} \geq 0.6Jy$) of the IRAS Point Source Catalogue. 
It has all 
the properties required for accurate calculation of peculiar velocities.
Selection from the IRAS Point Source Catalogue ensures near uniform 
selection across the sky. In particular, it avoids the problems with optically
selected surveys of matching the northern and southern skies. The PSC-z
survey also has excellent sky coverage, with $84.1\%$ of the 
sky in the survey. 
The flux limit of $0.6$Jy makes the PSC-z the deepest complete all sky 
IRAS redshift survey, 
ensuring that we have sufficient density of sampling to
accurately calculate peculiar velocities.     

The Least Action Principle is computationally expensive to apply, so we 
only use galaxies with $cz\leq 2000$\km (unless otherwise stated all of our
redshifts have been transformed to the Local Group frame using the
Yahil, Tammann \& Sandage et al (1977) correction). 
For an \ho of $80$ kms$^{-1}$Mpc$^{-1}$, our 
furthest distance of $17.8Mpc$ corresponds to  $cz=1420$kms$^{-1}$, so we 
have a reasonable buffer zone to the edge effects at $cz=2000$kms$^{-1}$. 
Galaxies with $cz\leq 500$kms$^{-1}$,
or within $12^{0}$ of M87 (ie in the Virgo cluster) are treated separately 
(see Sections \ref{locgal} and \ref{virgo} respectively). We impose a 
luminosity cut, to avoid any weighting problems with the poorly determined 
faint end of the luminosity function. 
The luminosity cut is such that all galaxies 
would have a flux of $0.6$Jy or greater when placed at $cz=1060$kms$^{-1}$.
We give each galaxy a weight $ w(cz)=1/\phi(cz)$, where $\phi(cz)$ is the 
 PSC-z selection function (Saunders et al 1999). The selection function
quantifies the loss of galaxies with distance, due to the flux cut. Its
value is the number density of galaxies that would be observed at 
a given redshift in the absence of galaxy fluctuations. 
The number of galaxies in this main sample is $929$.
This includes galaxies below the luminosity limit, as these may become 
sufficiently luminous when we iterate the positions to account for redshift 
distortions (see Section \ref{rediter}).
  
\subsection{Treatment of Masked Region}
\label{mask}

Although the PSC-z mask is small, we still have to account for it.
We do this by filling the sky behind the mask with fictitious galaxies.
To get a feel for the errors involved, we adopt two very different schemes
for doing this. The first is the ``random mask'', where galaxies are placed 
 at random in the masked region, with a space density consistent with the 
PSC-z selection function. The second is the ``cloned mask'' 
where we extrapolate
structures across the Galactic plane, from the galaxies we observe above
 and below the plane. Masked regions away from the Galactic plane (mainly 
the small areas where IRAS did not survey) are still filled with random 
galaxies. For the random mask we add a total of 169 ``galaxies'' and for the 
cloned mask we add 145 ``galaxies''.

\subsection{Grouping of the Galaxies}
\label{group}

As we described in Section \ref{nam}, we are looking for 
minimum solutions to equation \ref{dact},
as the global minimum is the solution closest to the Hubble flow. This 
solution is likely to be adequate, except in regions at the centres of groups
and clusters. In these regions, the motions of galaxies are likely to 
be a general stationary point of the action, as the orbits will have   
crossed (see P89, P90). So to avoid problems with these stationary solutions,
we group our galaxies and just follow the motion of the centre of mass.
The grouping of galaxies also helps us to deal with redshift distortions.
We attempt to iterate out the coherent component to the redshift 
distortions during our calculations (see Section \ref{rediter}). However
there is nothing we can do about the incoherent component to redshift
distortions generated on small scales. By grouping the galaxies 
we can hope to average out the worst effects of these incoherent redshift
distortions.
To group we use a ``Friends of Friends'' algorithm, similar to Huchra \& Geller
(1982). The linking is 
$ \Delta d=0.6 $\mpcs  tangentially, and $ \Delta z = 150$\km  radially,
for galaxies with average redshift inside $ cz=1060$kms$^{-1}$. 
For galaxies outside the volume limited sample we scale the lengths, to allow 
for the decreasing density of selected galaxies. This is done according to,
\begin{eqnarray}
\Delta d & = & 0.6 \left( \frac{ \phi (1060)}{ \phi(cz)}
  \right)^{\frac{1}{3}} h^{-1}Mpc, \\
\Delta z & = & 150 \left( \frac{ \phi (1060)}{ \phi(cz)}
  \right)^{\frac{1}{3}} kms^{-1}. 
\end{eqnarray}
The radial linking is so much greater to allow for the effects of 
velocity dispersion inside the groups.
Once a galaxy is assigned to a group, it is considered to be at the
centre of mass of that group.   
We have compared the groups with those of the 
Tully (1987) catalogue. Most of the 
time the two agree fairly well, though there were differences, but we are 
confident that the grouping is working well.
 
\subsection{ The Virgo Cluster}
\label{virgo}

The Virgo Cluster is the most important object for the dynamics of the 
local volume, so it is important to treat it well in our calculations. 
To do this, we have treated it separately from the rest of the galaxies. 
This is necessitated by the large size and velocity dispersion of the cluster.
The linking lengths of the grouping algorithm, have been chosen for small to 
medium groups, and are thus inappropriate for the Virgo cluster. We also 
need to take account of blueshifted galaxies, which outside of the 
Local Group, are only present in the Virgo Cluster. We consider all galaxies
within $12^{\circ}$ of the Virgo core (which we take to be M87), to be a
part of the Virgo cluster.
Provided that they have a redshift below 
$cz=2000$\km, galaxies with blueshifts in this region of the sky are also 
placed in the Virgo cluster. We cannot extend the upper redshift limit beyond
$cz=2000$\km due to contamination from background galaxies.
On these criteria, $105$ galaxies are placed in 
the Virgo Cluster, which we place at the angular position of M87. For its
redshift we use the value given by Bingelli et al (1987), not including
the dwarf galaxies.   
  
\subsection{ Tidal Effects}

Galaxies outside $cz=2000$\km, could potentially generate large tidal 
motions. We thus calculate these tidal motions using linear theory and 
add them to the Least Action Principle results. We use the standard equation,
\begin{equation}
{\bf V}=\frac{H_{\circ}\beta}{4 \pi} \Sigma_{i} w_{i} 
\frac{{\bf r}_{i}}{r^{3}}.
\end{equation}
Where $ w_{i} $ is the weight of the $i$th galaxy (see Section \ref{pscz}). 
The tidal velocities are calculated with all galaxies between
$ cz=2000$\km to $ cz=15000$\km in the PSC-z. 
The galaxies are placed in the Local Group frame if they are inside 
$cz=3000$\km, but in the CMB frame if they are outside.
 We use a random mask model throughout for the tidal effects.
The tidal velocities are also
added to the peculiar velocities when we compensate for redshift distortions.
Clearly the positions of our galaxies, with just the switch in frames at
$cz=3000$\km are very crude. We have compared the tidal velocities calculated 
using these positions, with those obtained using a full reconstruction of 
real space positions. The results compare very well, with most velocities
differing by less than $10$\km.   

\subsection{Redshift Distortions}
\label{rediter}

The redshifts of the galaxies and groups will suffer from redshift
distortions due to their peculiar motions. The simple use of these redshifts
as distances would cause large errors. As a result we attempt to iterate out
the effects of redshift distortions in a manner similar to S95. This involves
iterating the position of a galaxy/group so that the quantity 
$1/c(H_{\circ}d+v)$, is equal to the redshift of the galaxy
(where $d$ is the distance to the galaxy and $v$ its radial 
peculiar velocity, relative to the Local Group). The peculiar velocities 
are calculated using the Least Action Principle (see Section \ref{nitty}).
We iterate the positions $10$ times, and the distance a galaxy 
is allowed to move  is 
slowly scaled down throughout the $10$ iterations. By the last iteration
most of the residues are small ( $ \leq 0.5$\mpcs  in most cases). 

We have proper distances for most of the local galaxies (ie those with 
$ cz \leq 500$\km). So, in order to avoid the vagaries of iterating
out the redshift distortions we place these galaxies, where we can,
at our distance estimates. It is very important to place these galaxies 
at the correct distances, as due to their close proximity, they can 
potentially have a 
large effect on the motion of the Local Group. The motion of the Local 
Group is crucial as all velocities are measured relative to it.    
To be consistent we also place the galaxies for which we have Cepheid 
distances outside $cz=500$ \km, at those Cepheid distances.
An exception is Virgo, which is allowed to iterate to near its distance, 
and then placed at its distance. This is to allow the correct iteration
of the redshifts of galaxies near the Virgo Cluster. 
Although the grouping is done with the 
galaxies in the Local Group frame, we then attach galaxies below the luminosity
cut to groups, or treat them as massless particles. Then, as we iterate out 
the redshift distortions, and change the distance to galaxies, 
we re-assess whether galaxies are above or below the luminosity limit and act
accordingly. This may seem quite a trivial effect, but a substantial number
of galaxies do change status, in a very systematic way across the sky.

\subsection{ Weighting of Local Galaxies}
\label{locgal}

The galaxies inside $ cz=500 $\km  are treated differently from the rest
 of the survey. This is partly to reflect our increased knowledge of 
these galaxies and partly due to their importance in determining the Local
Group's velocity (relative to which all other velocities are observed).
We did not apply our grouping algorithm to these galaxies, but instead use
the 7 groups defined by Kraan-Korteweg \& Tammann (1979), but follow
the pleading of Schmidt \& Boller (1992b) for a subdivision of the B7 group.

\begin{table}
\begin{tabular}{l|c|c}
group/galaxy & PSC-z Weight & Optical weight \\
             &         &      \\
Local Group  & 0.09    & 0.15  \\
B1(Maffei)   & 0.06    & 0.23  \\
B2(M81)      & 0.15    & 0.09  \\
B3(M101)     & 0.09    & 0.10  \\
B4           & 0.03    & 0.02  \\
B5(IC4182)   & 0.06    & 0.04  \\
B6(N5128)    & 0.15    & 0.17  \\
B7a(N300)    & 0.03    & 0.02  \\
B7b(N253)    & 0.06    & 0.05  \\
N3621        & 0.03    & 0.02  \\
N3115        & 0.00    & 0.04  \\
N2903        & 0.03    & 0.02  \\
N6946        & 0.03    & 0.04  \\
N6503        & 0.03    & 0.00  \\
N4605        & 0.03    & 0.00  \\
N2683        & 0.03    & 0.00  \\
N3593        & 0.03    & 0.00  \\
N0625        & 0.03    & 0.00  \\
N1313        & 0.03    & 0.00  \\
\end{tabular}
\caption{PSC-z and Optical weightings for the local galaxies. 
The total weight in each column is $1.00$.
The group names refer to those of Kraan-Korteweg \& Tammann (1979). We give
alternatives names in brackets, 
normally the brightest galaxy. \label{massmodel} } 
\end{table}

We have two distinct ways of assigning weights to these local galaxies.
The first is  using the PSC-z catalogue, each group or galaxy is given 
a weight according to how many galaxies it contains above the luminosity 
limit; when placed at its distance for $ H_{0}= 80 $\km Mpc$^{-1}$.
The actual value of \ho 
used makes little difference to the weights. 
If a galaxy has no distance and is not in a group with
a distance, we use its redshift. 
 A second weighting is given by optical luminosity.
We take the optical luminosity data from the catalogue of local galaxies
compiled by Schmidt \& Boller (1992a), each galaxy and group is 
weighted in proportion to its luminosity and and all galaxies and groups
with a luminosity above $ 10^{10} L_{\odot} $ are included, (B7a is just
under this limit, but too close to the Local Group to cut). To tie these 
relative weights into the PSC-z weightings used for the rest of the galaxies
we multiply the relative weights by the total weight of PSC-z galaxies
inside $ cz=500 $\km for $ H_{0}=80$ kms$^{-1}$Mpc$^{-1}$.
The results of these schemes can be seen in Table \ref{massmodel}.
The main difference in the two weighting schemes, is that while the PSC-z
weighting tends to be fairly even across the galaxies and groups, 
the optical weighting tends to favour a few large groups. With the two 
possible mask compensations and the two possible local weightings,
 we have four possible catalogues.

We have two main aims in having two weighting schemes. Firstly, we
wanted to choose the weighting scheme which optimised the results, and
secondly, to see
how robust the results are to different weighting schemes. The issue of 
robustness is important. 
Although either of these weighting schemes might
be on average a good tracer of the total mass, 
we expect that the weight assigned to any one galaxy or group
is subject to large errors. We can gauge
the possible effects of these errors by looking at the 
results of the two weighting schemes.

\subsection{Distances}
\label{dist}

We present our distances in Table \ref{distances}. All the distances are
derived from Cepheids and seven of the 12 are inside $10Mpc$, making our
analysis sensitive to small scales. All of the galaxies are in the 
PSC-z catalogue. The errors consist 
of the quoted distance errors given in the references added in quadrature
to a ``projection`` error for groups. The projection error accounts for the 
fact that the galaxy to which we have a distance may not be at the centre 
of the mass of the group. This error is given by
\begin{equation}
\Delta d = \frac{d}{ \sqrt{2} n } \sum_{i} \Delta \Theta_{i}. 
\end{equation}
Where the $ \Delta \Theta_{i} $ are the angular distances of the galaxies from 
the centre of mass, $n$ is the number of galaxies in the group 
and $ d $ is the distance to the group. This error is 
normally the minor partner, only dominating for B5, B6 and B7a.
It is possible that our distances are affected by systematic errors.
The most likely sources are either problems with the calibration due to 
errors in the distance to the LMC, or errors due to metallicity effects.
Problems with the calibration at the LMC will affect all the distances 
by the same fractional amount. Thus, although it will have direct effect
on the estimate of \hoc, it will leave the rest of our conclusions unchanged.
Significant metallicity effects would have the potential to effect all of our 
results, but luckily they are thought to be small (Kennicutt et al 1998).

To allow us to better combat redshift distortions in the local volume, 
we place the groups B1, B4 and B7b and the galaxy N3115 
at non-Cepheid distance measures.
To avoid possible systematic errors these are not used in the 
peculiar velocity comparison. B1 is placed
at $4.2Mpc$, from the K-Band surface brightness fluctuations measured by
Luppino \& Tonry (1993). Graham et al (1982) places B7b $1Mpc$ more 
distant than B7a, so we place it at $3.1Mpc$. The Tully-Fisher ratios of 
B4 and B5 from Willick et al (1997b), led us to place B4 at $3.7Mpc$.
N3115 is placed at $10.8Mpc$ from the Planetary Nebula Luminosity Function 
determination of Ciardullo et al (1993).  

\begin{table}
\begin{tabular}{l|l|l|l|l}
galaxy and & dist     & error &  red-     & reference \\
group      & $ Mpc $&  $ Mpc $& shift &            \\
           &          &           &       &           \\
M81 (B2)   & 3.6      &  0.45 & 192       & Freedman et al 94 \\
M101 (B3)  & 7.4      & 0.7   & 374       & Kelson et al 96 \\
IC 4182 (B5) & 4.7      & 0.5   & 385     & Saha et al 94  \\
N5253 (B6)  & 4.1     &   0.55 & 260      & Saha et al 95  \\
N300 (B7a)  & 2.1     & 0.3    & 137      & Freedman et al 92 \\
N3621       & 6.8     & 0.6    & 435      & Rawson et al 97    \\
N0925 (N1023) & 9.3   & 0.8   & 807     & Silbermann et al 96 \\
N3351 (LEO I)& 10.8  &  0.9     & 625     & Graham et al 97 \\
N3368 (LEO I)  & \multicolumn{3}{l}{ Average Used} & \& Tanvir et al 95 \\
N2090       & 12.3  &  0.9     & 753     & Phelps et al 98     \\
N2541       & 12.4  &  0.7     & 593     & Ferrrase et al 98 \\
N7331       & 15.1    & 1.4   & 1115    & Hughes et al 98 \\
Virgo       & 17.8  &  1.8    & 960     & Freedman et al 98 \\
\end{tabular}
\caption{A Table of distances to nearby galaxies and groups, the redshifts
are in the Local Group frame with the Yahil et al (1977) correction.  
\label{distances} }
\end{table}

\section{Results}
\label{resvel}

In this Section we discuss the results from our velocity comparison.
The first Section \ref{nitty}, deals with the exact details of how we 
apply the Least Action Principle. We then discuss the goodness of fit in
Section \ref{gof} and the confidence limits we obtain 
on \omeg and \ho in Section \ref{cl}.

\subsection{Application of Least Action Principle}
\label{nitty}

As explained in Section \ref{nam}, we iterate the $ C_{i,n}^{ \alpha} $
towards minimum using equation \ref{cit}. 
A crucial point is how many iterations are required 
to achieve good convergence. We have approached this problem in a 
different manner to previous papers (P90, S95), who looked at the changing 
value of the gradient and stopped calculations when this was 
sufficiently small. Instead we compared the velocities predicted after
a number of iterations, with those obtained after $1000$ iterations.
 Typical results can be seen in Table \ref{converge}.
\begin{table}
\begin{tabular}{l|cccc}
Number of & \multicolumn{4}{c}{Accuracy of velocities ($kms^{-1}$)}  \\
iterations & $\leq5$ & $\leq10$ & $\leq20$ 
& $\leq50$  \\
10  & 2\%   & 17\%  & 64\% &  97\% \\
20  & 1\%   & 10\%  & 66\% &  99\% \\
50  & 0\%   &  6\%  & 63\% &  99\% \\
100 & 5\%   & 60\%  & 94\% & 100\% \\
150 & 60\%  & 92\%  & 98\% & 100\% \\   
250 & 97\%  & 98\%  & 99\% & 100\% \\
\end{tabular}
\caption{The convergence of velocities with number of iterations. The 
velocities are compared to those obtained with $1000$ iterations. The results
here are for $\Omega_{0}=0.3$, $H_{0}=70kms^{-1}Mpc^{-1}$,
PSC-z local weighting and  cloned mask. Results from other models 
are very similar. \label{converge}}
\end{table}
As the table shows, initial convergence to rough values is very quick,
but final convergence takes a great many iterations; with velocities only
changing by a small amount each iteration. We thus choose to iterate 
the $ C_{i,n}^{ \alpha} $ $250$ times, iterating galaxy positions every $10$
 $ C_{i,n}^{ \alpha} $ iterations for the first 
$100$ iterations (see Section \ref{rediter}). 
The rough velocities
 used to iterate the positions are easily adequate; as we only aim to 
drive the residues to $ \leq50kms^{-1} $.

In integrating equation \ref{cit}, we use $10$ timesteps. Comparisons with 
results calculated with $40$ and $100$ timesteps, suggests that this gives us
$ \sim85\%$ of velocities within $20 kms^{-1}$ and $ \sim98\%$ of velocities
within $50 kms^{-1}$. As convergence is slightly slower 
for a larger number of timesteps,
it is clearly preferable to have (as we do) a low number of timesteps
combined with a large number of iterations, (the computational time
is directly proportional to the number of timesteps multiplied by the number
of iterations).
   
As mentioned in Section \ref{nam}, the solution for which we seek is the
global minimum. This should give us the galaxy orbits which are closest 
to the Hubble flow, and is the natural extension of linear theory. The
existence of a global minimum is guaranteed as $S \rightarrow \infty$ as
$ C_{i,n}^{ \alpha} \rightarrow \infty $. However, it is also possible that 
other local minimum exist. Indeed trials from large random values of 
$ C_{i,n}^{ \alpha} $ reveal that multiple minima exist. As a result 
approximately $15$ \% of galaxies and groups can have two distinct solutions
(by  distinct solution we mean a difference in velocities of greater than
$50$ \km). Unfortunately, it is impossible to calculate the differences 
in the value of the action between solutions accurately enough to 
ascertain which solution corresponds to the global minimum. 
For pure Hubble flow we would have $C_{i,n}^{ \alpha}=0$, 
so it seems reasonable that the solution obtained when we iterate from 
$C_{i,n}^{ \alpha}=0$ is the global minimum. Certainly, the use of this 
solution has become conventional in Least Action work (ie S95). 
Thus it is this solution 
which we use for the results in Sections \ref{gof} and \ref{cl}.
A particular problem we experience is that the Local Group has two distinct 
solutions for three of the catalogues. Only for the optical local weighting 
and random mask catalogue do the trials from large random values of 
$C_{i,n}^{ \alpha}$ fail to reveal a second solution. For the other three
catalogues, the two Local Group solutions have the potential to cause great 
problems as all velocities are measured relative to the Local Group. 
We have, by fixing the initial values for $ C_{i,n}^{ \alpha} $, 
explored the other Local Group solutions for the three catalogues. 
The results found with these solutions are consistent 
with those presented below, with similar confidence limits being 
obtained. For example with the PSC-z local weighting and cloned mask we 
obtain limits $0.1 \leq \Omega_{0} \leq 0.40$ and  
$ 65 \leq H_{0} \leq 72.5$ \km Mpc$^{-1}$ with the ``standard'' solution, 
iterating from $ C_{i,n}^{ \alpha}=0 $ and limits of 
$0.05 \leq \Omega_{0} \leq 0.25$ and  
$ 65 \leq H_{0} \leq 70$ \km Mpc$^{-1}$ with the ``second'' solution.

\subsection{ Goodness of Fit}
\label{gof}

A qualitative impression of the goodness of fit can be obtained from Figure
\ref{scath}. This presents scatterplot diagrams of the predicted velocities
from the Least Action Principle calculations against the observed velocities
from the Cepheid distances. All of the plots show results from calculations 
with a PSC-z local weighting and Cloned mask. The error bars are given by 
$ \sigma_{oi}=H_{0} \Delta D $. Where $\Delta D$ is the error in the distance
and $ \sigma_{oi}$ is the resulting error in the observed velocity. For all
our results we have calculated the observed velocity errors using 
$ H_{0}=70kms^{-1}Mpc^{-1}$; we take no account of errors in the predicted
velocities. For all the scatterplots where \omeg and \ho are near best fit,
(ie around $ \Omega_{0}=0.2$, $H_{0}=70 kms^{-1}Mpc^{-1}$, see below)
almost all of the points lie on the 
observed velocity equals predicted velocity line,
 to within their errors. The only exception is the $N1023$ group
(marked in the bottom left plot), even at best fit this point is a 
considerable distance from the line. So with the exception of this galaxy 
we are getting a very good fit. 

\begin{figure*}
\begin{center}
\resizebox{5.5cm}{!}{\includegraphics{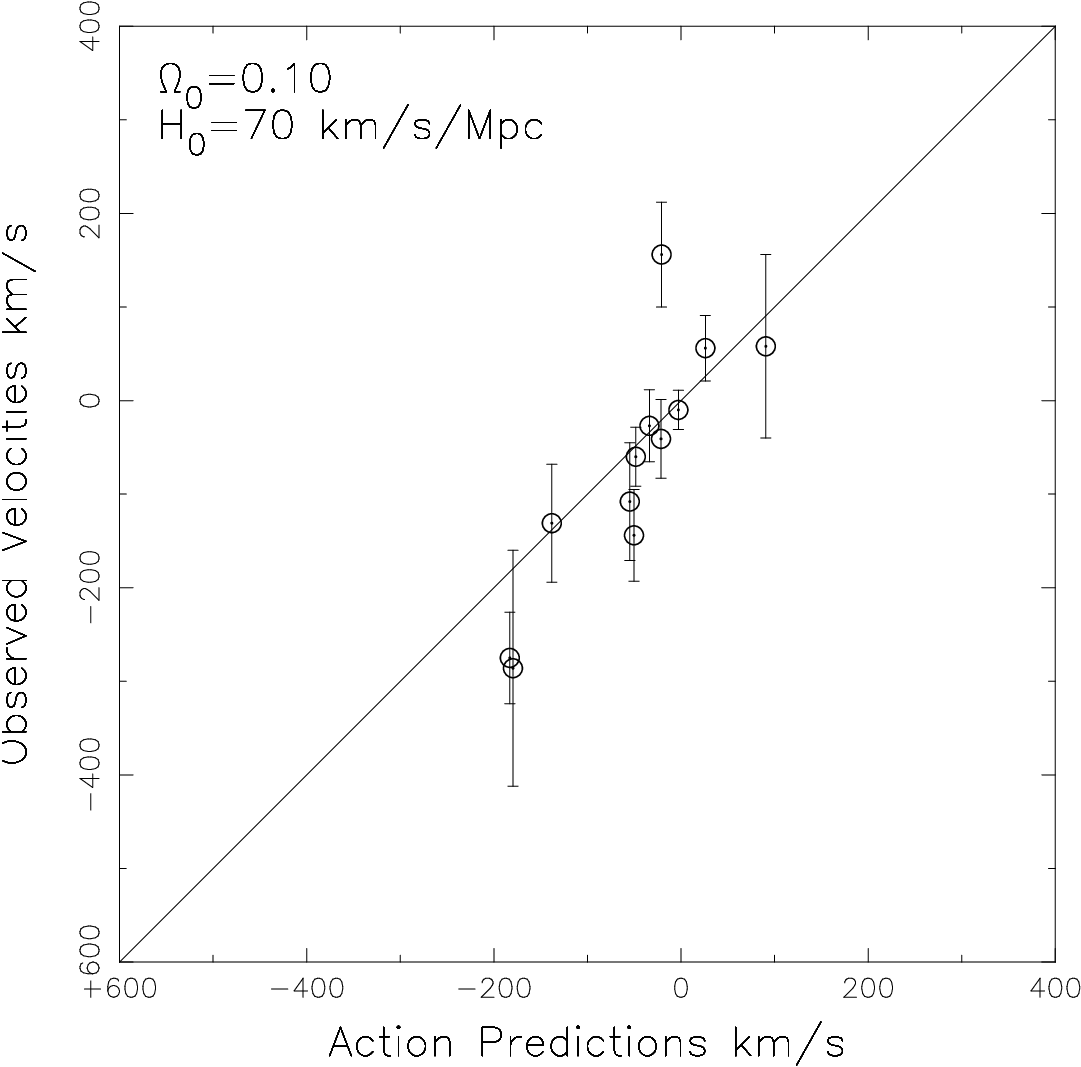}}
\resizebox{5.5cm}{!}{\includegraphics{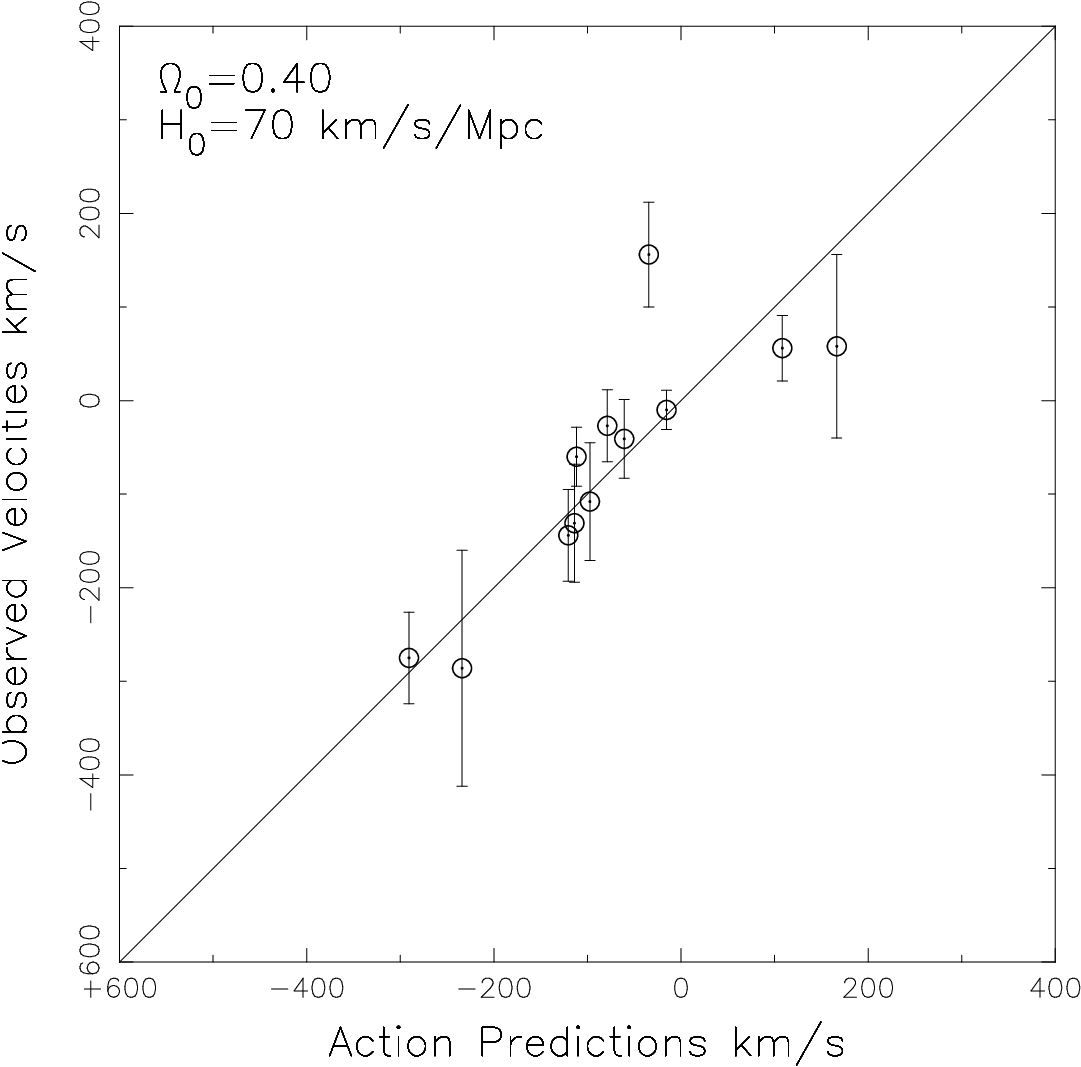}}
\resizebox{5.5cm}{!}{\includegraphics{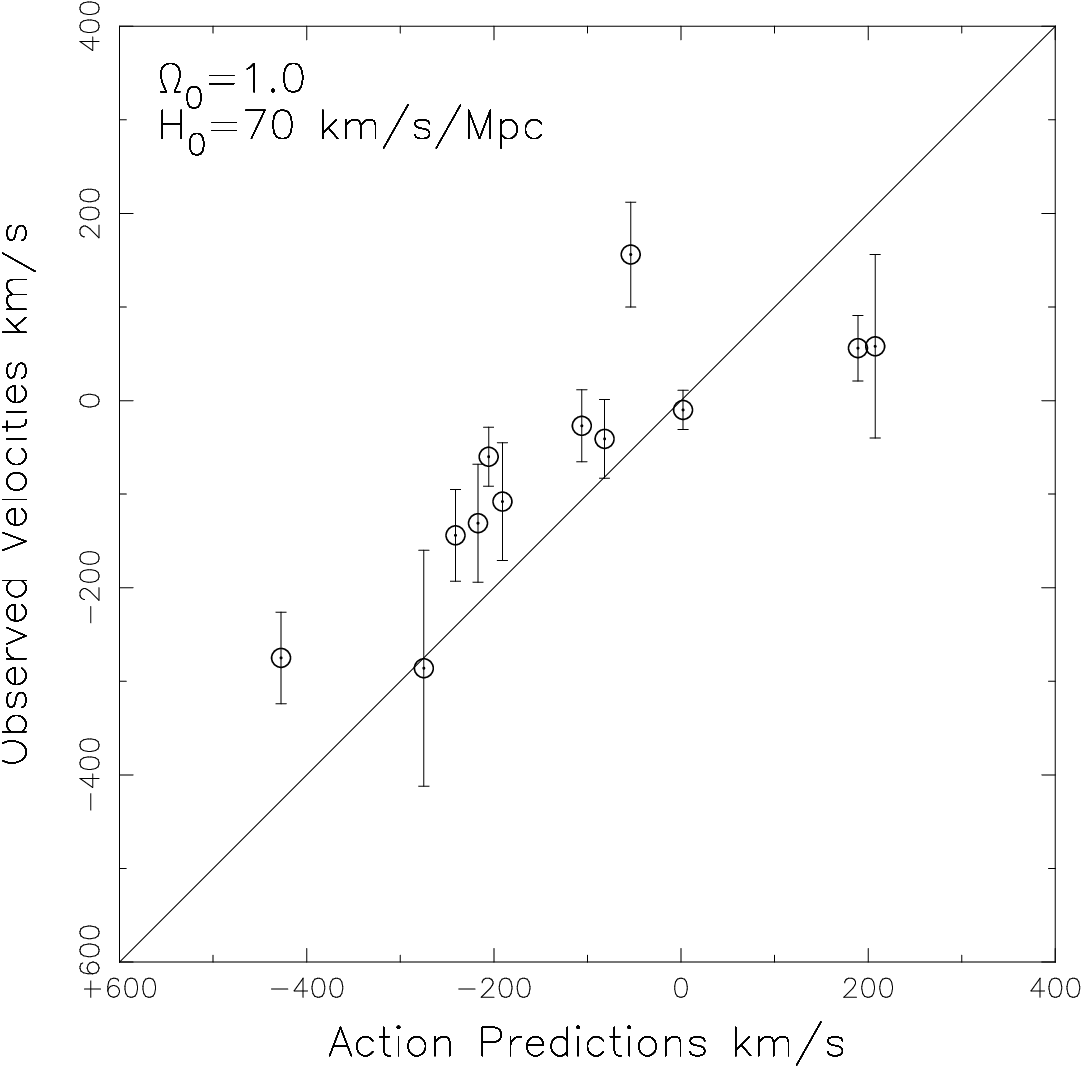}}
\resizebox{5.5cm}{!}{\includegraphics{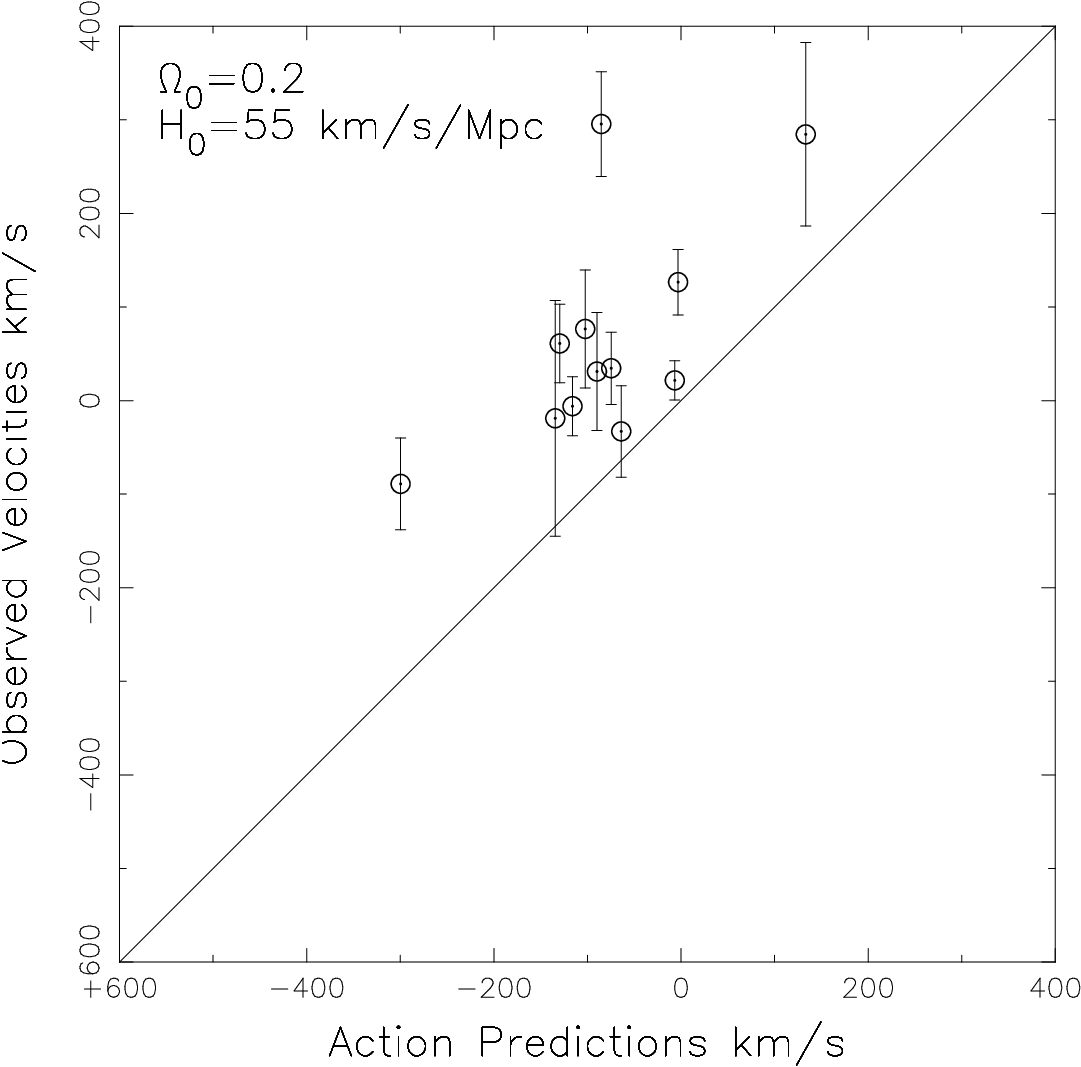}}
\resizebox{5.5cm}{!}{\includegraphics{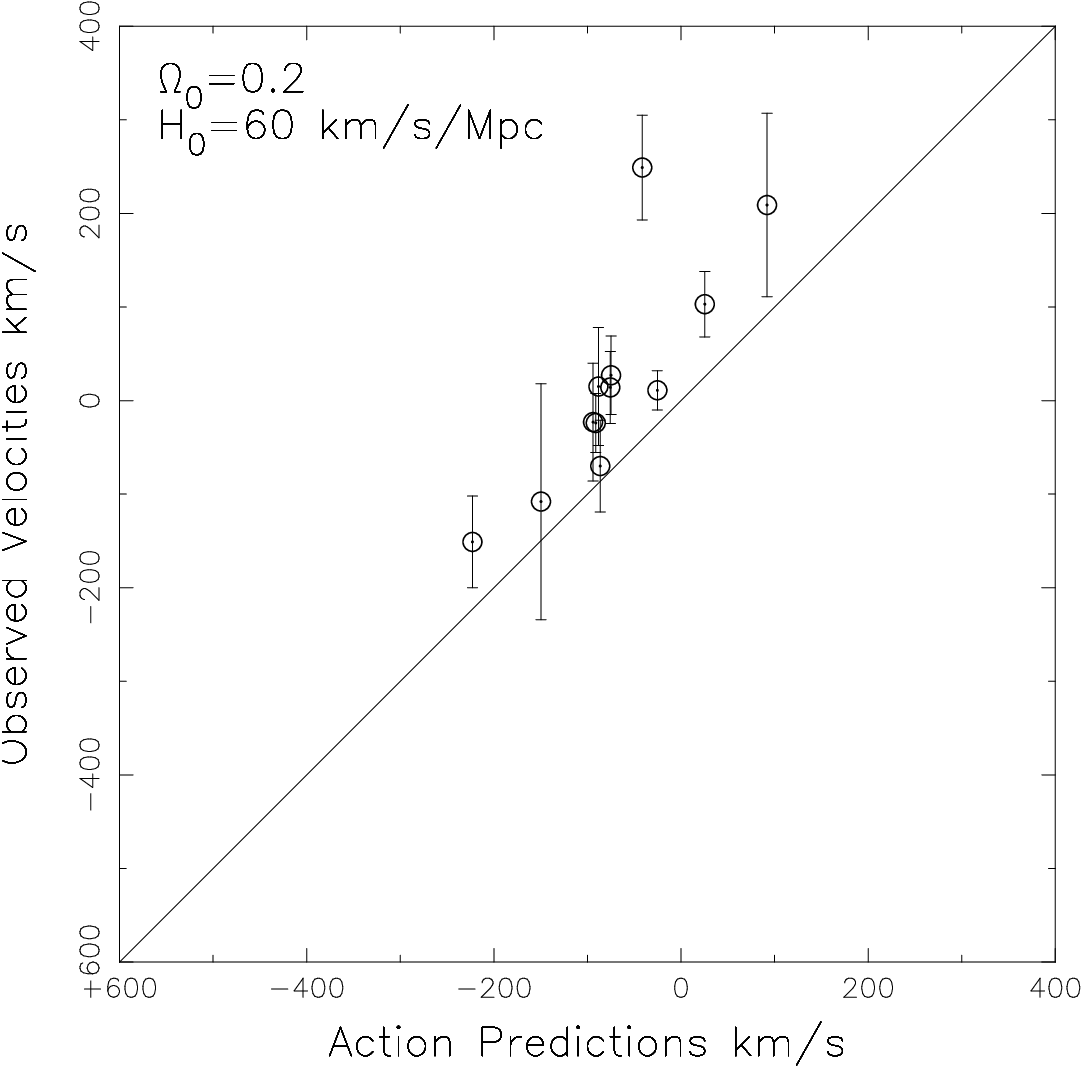}}
\resizebox{5.5cm}{!}{\includegraphics{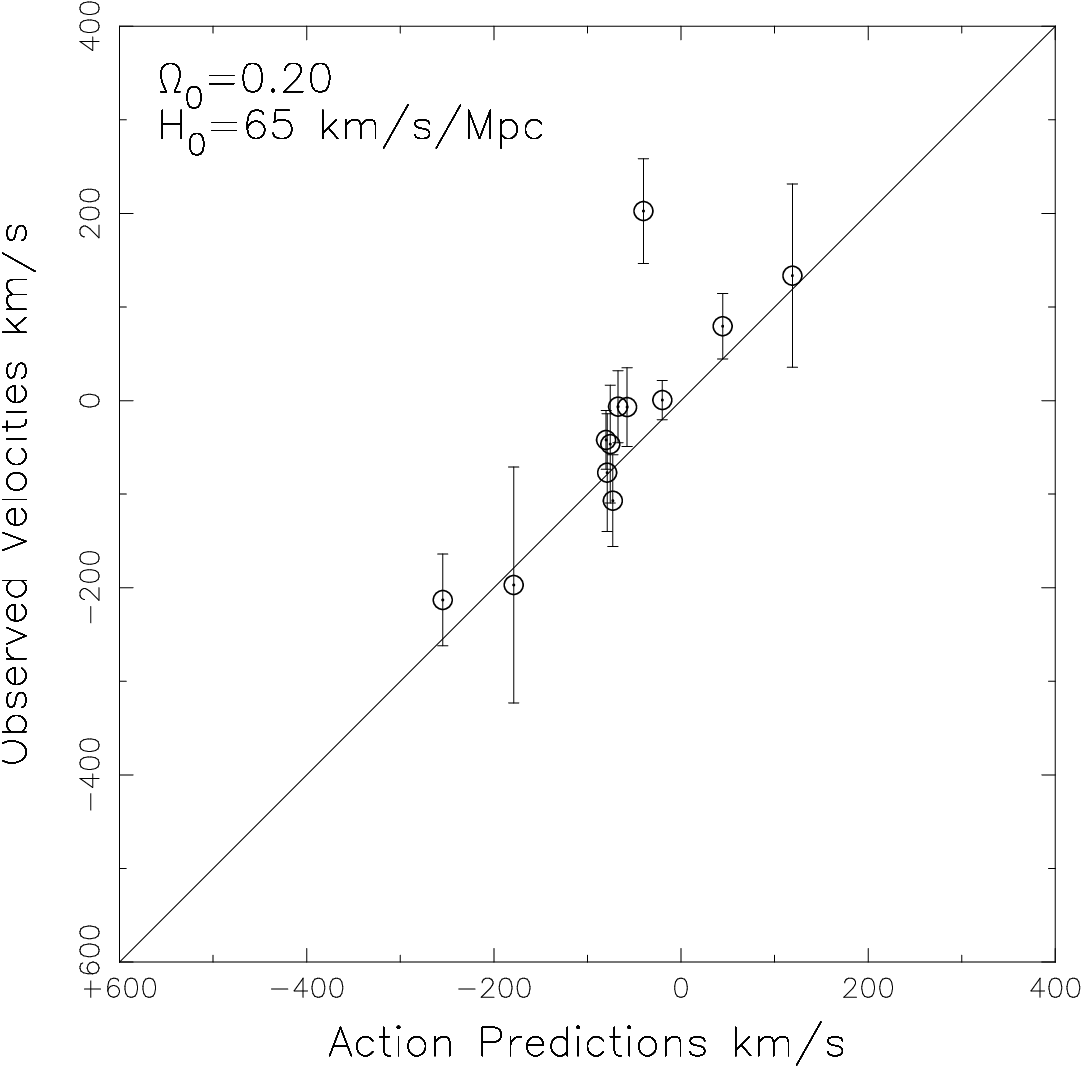}}
\resizebox{5.5cm}{!}{\includegraphics{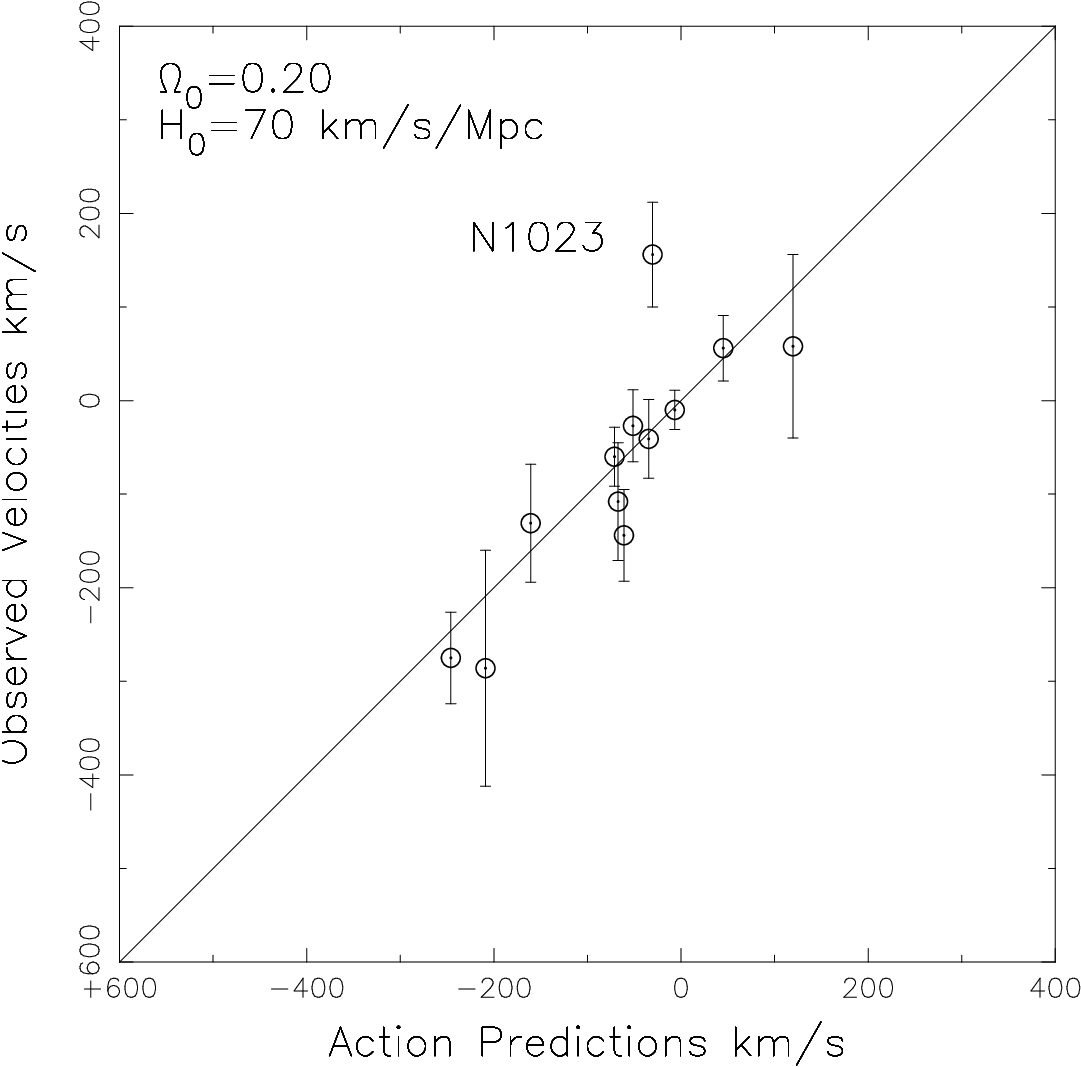}}
\resizebox{5.5cm}{!}{\includegraphics{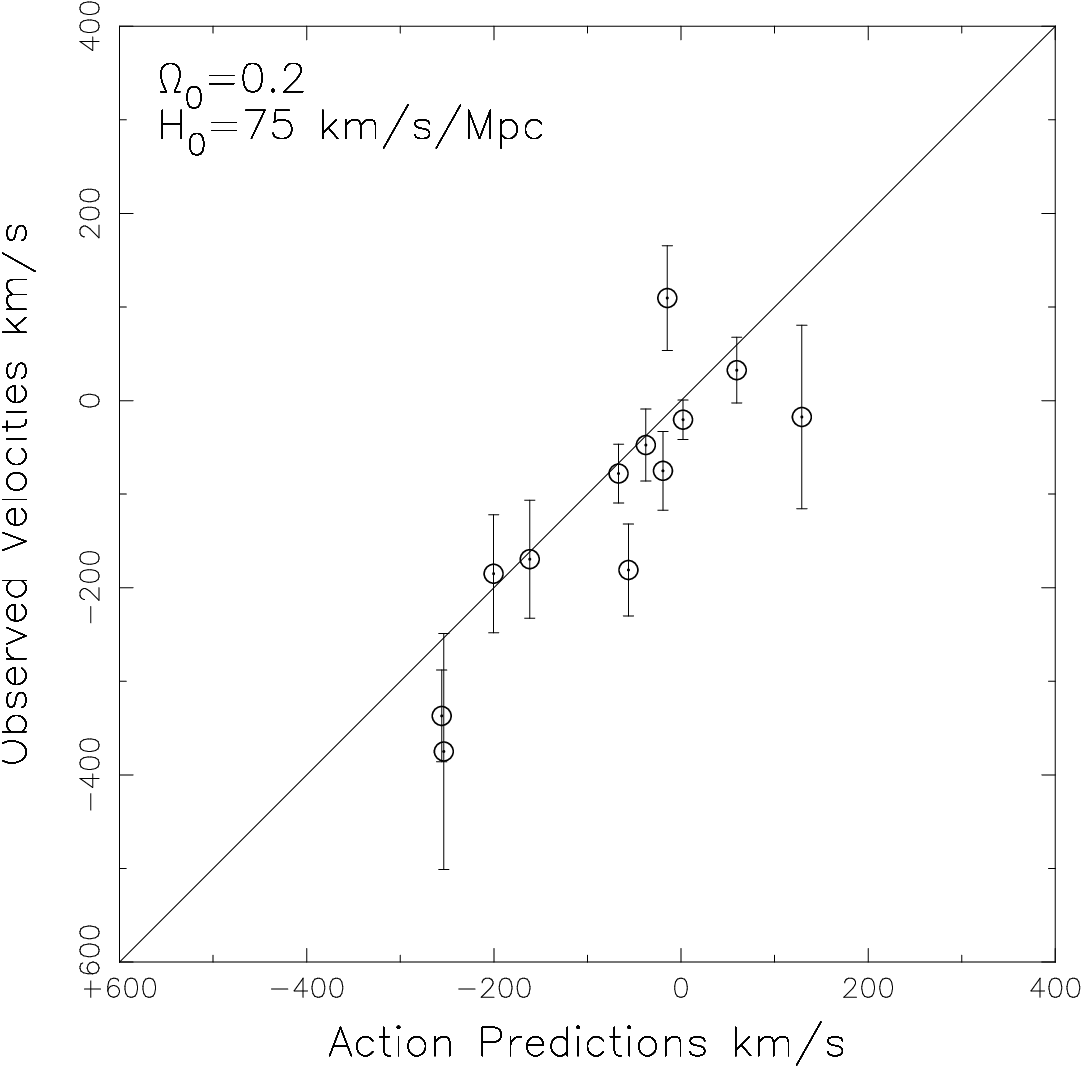}}
\resizebox{5.5cm}{!}{\includegraphics{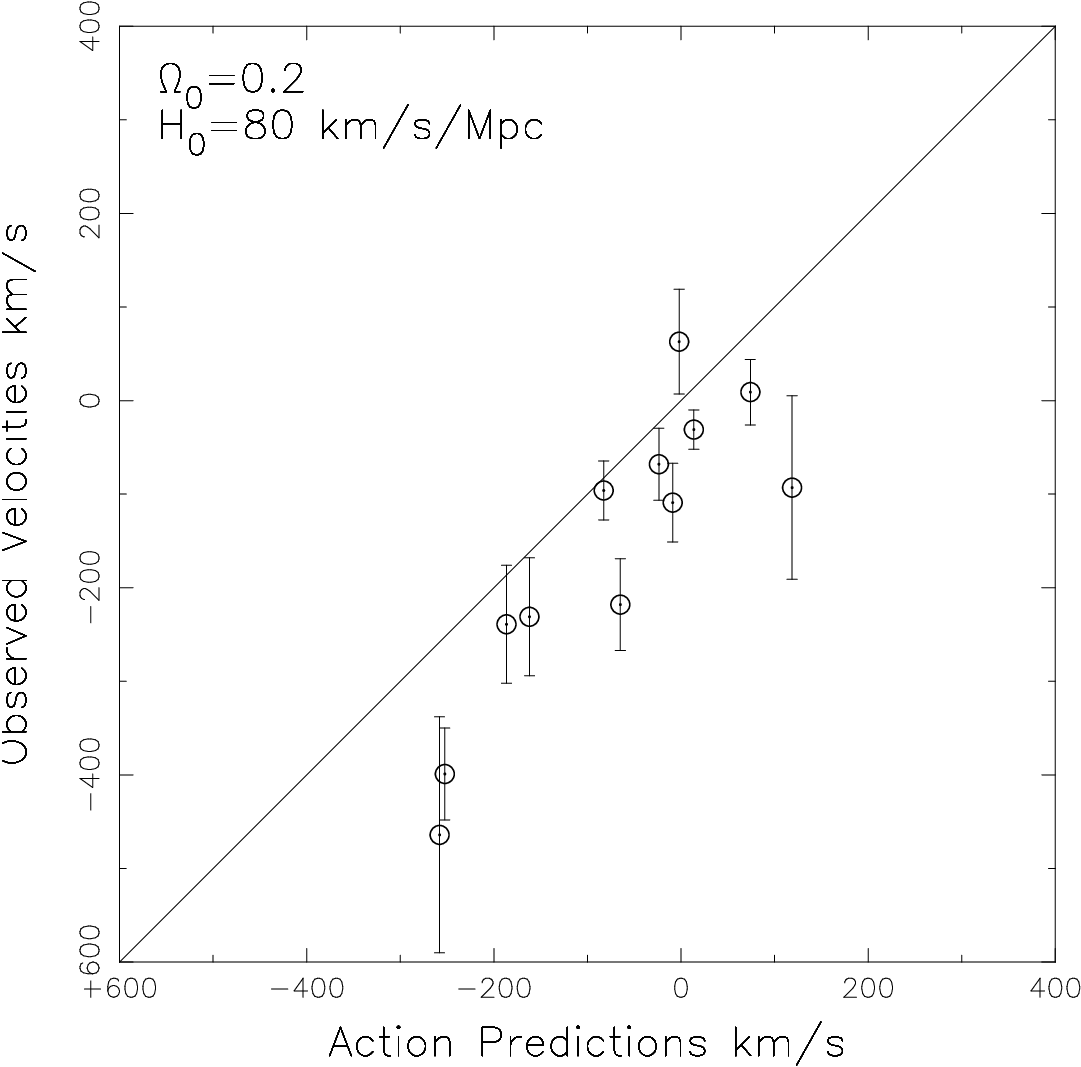}}
\caption{Scatterplots showing the effect of varying\ho and \omegc. 
All results are with PSC-z local weighting and a cloned mask model. The line
shown is observed velocity= predicted velocity. 
\label{scath} }
\end{center}
\end{figure*}

To make our results more quantitative we performed a reduced \ci test,
\begin{equation}
\chi^{2} = \frac{1}{N-M} \sum_{i} \frac{ (v_{pi}-v_{oi})^{2}}
{ \sigma_{oi}^{2}}.
\end{equation}
Where $ v_{pi} $ is the predicted velocity from the Least Action Principle, 
$ v_{oi} $ is the observed velocity and $ \sigma_{oi} $ its error.  
N is the number of data points and M the number of fitted parameters.

Predicted velocities using the Least Action Principle where obtained
on a grid of \omeg and \ho for the four possible input catalogues.
 In Table \ref{goodfit} we present \omegc, \ho 
and the value of the reduced \ci statistic at best fit for the four 
catalogues. We do this using all the galaxies and excluding the outlier 
N1023 group.

\begin{table*}
\begin{tabular}{l|l|l|l|l|l|l|l}
Mask         & Local      & 
\multicolumn{3}{l}{Full Sample}   & \multicolumn{3}{l}{Excluding  N1023}    \\
compensation & Weightings &  \ci & \omeg & \ho & \ci & \omeg & \ho   \\
random       & PSC-z      & 1.28 & 0.40 & 75 & 0.64 & 0.30 & 70    \\
cloned       & PSC-z      & 1.56 & 0.25 & 70 & 0.46 & 0.25 & 70  \\
random       & optical    & 2.41 & 0.25 & 72.5 & 1.11 & 0.20 & 70  \\
cloned       & optical    & 2.27 & 0.15 & 72.5 & 0.76 & 0.10 & 67.5   \\
\end{tabular}
\caption{ Best Fit \ci values, with and without N1023. The values
of \omeg and \ho given are those at the best fit \ci.   \label{goodfit} }
\end{table*}

For the reduced \ci test we are performing the $95 \%$ confidence limit
is $1.83$. Thus the results confirm the impression of the scatterplots
that the fit is very good, especially when the $N1023$ group is discarded.
The PSC-z local weight is performing better than the optical weight and by
a smaller margin the cloned mask is preferred to the random mask.
Importantly we see that the results are reasonably robust to both 
local weighting and mask compensation, with the best fit \omeg and \ho 
being relatively close for the four models.  

We are not sure why we are having such problems with the $N1023$ group.
It is not in the $15\%$ of galaxies and groups for which trials from 
large random values of $C_{i,n}^{ \alpha}$ reveal multiple solutions.
Nor is the predicted peculiar velocity strongly effected by the multiple 
solutions of the Local Group.
It lies reasonable close to the galactic plane, so we could be experiencing 
problems in how we compensate for the mask. The group is an outlier for 
both the mask models, but it is possible that both models miss an important 
structure (ie a reasonable compact group) that is essential to accurately 
predict the velocity of $N1023$. Another possibility is a problem with the
distance to the group. This seems less likely as the 
Cepheid distance agrees well 
with other measurements of $9.6Mpc$ from the Surface Brightness Fluctuation
 method (Tonry et al 1997), and $10.1Mpc$ from the Planetary Nebula Luminosity 
Function method (Ciardullo et al 1991). For the purposes of the next Section,
where we calculate confidence intervals, we chose to exclude the $N1023$ group.
If either of the problems mentioned above are true, then clearly we are well 
justified in doing so. However, even if $N1023$ is a natural outlier then
it is necessary to exclude it. The small number of galaxies that we have are 
not robust to its effect and fairer limits are obtained without it.

\subsection{ Confidence Intervals}
\label{cl}

The velocities that we predict for galaxies and groups depend mainly an the 
input value of \omegc. There is also a weak dependance on \ho as this changes
where we place the galaxies with measured distances.
The main effect of 
\ho in the peculiar velocity comparison, is that it changes the observed 
velocities (given by $v=cz-H_{0}d $) not the predicted velocities. 
Figure \ref{scath} shows the effects changing of \omeg and \hoc.
Changing \omeg stretches or compresses the range of predicted velocities, 
thus for low \omeg our range of predicted velocities
 is too narrow, whilst for higher \omeg it is too broad.
 Changing \ho shifts the observed velocities up and down, thus for low \ho
 the observed velocities tend to be too large, whilst for 
high \ho  they tend to be small. In effect we can think of \omeg as changing 
the gradient of our data and \ho as changing the zero-point, only for 
the certain values of \omeg and \ho will our data be consistent with the
observed velocity equals predicted velocity line.    

We can make a quantitative assessment of the 90 \% confidence limits 
by using a 
$ \Delta \chi^{2}$ test. The results of this is shown in Table \ref{confid}.
For all four models, we get similar ranges of \omeg and $H_{0}$. 
This indicates
that we are robust to the Local weighting scheme and to the errors we expect 
in weighting given to any individual galaxy or group. We are also reasonable
robust to mask compensation model, though the random mask does allow for a 
higher upper limit on $\Omega_{0}$. We take our final confidence limits from 
the union of the results with the PSC-z local weighting, as this weighting 
was preferred. This gives $ 65 \leq H_{0} \leq 75$ \km Mpc$^{-1}$,
and $ 0.1 \leq \Omega_{0} \leq 0.60$. We can convert to $ \beta=
\Omega_{0}^{0.6} / b $, 
under the assumption that $b=1$ (see Subsection \ref{subias}).
 This gives us $ 0.25 \leq \beta \leq 0.75 $.
As in the determination of \ho the distance to Virgo is somewhat controversial
we have calculated an \ho confidence interval without using this distance
for the PSC-z local weighting and cloned mask catalogue.
We get the same range as before 
(ie  $ 65 \leq H_{0} \leq 72.5$ \km Mpc$^{-1}$), so our estimate does not 
depend very heavily on the distance to Virgo. For interest, we tend to 
calculate an infall into Virgo of around $200$kms$^{-1}$, for $ \beta=0.5$. 

\begin{table}
\begin{tabular}{l|l|lr|lr|c|lr}
Mask         & Local      & \ho Range & \omeg range \\
compensation & Weightings & $ kms^{-1}Mpc^{-1} $ &    \\
random       & PSC-z      & 67.5-  75.0 & 0.10- 0.60   \\
cloned       & PSC-z      & 65.0-  72.5 & 0.10- 0.40    \\
random       & optical    & 65.0-  72.5 & 0.05- 0.40   \\
cloned       & optical    & 65.0-  72.5 & 0.05- 0.30   \\
\end{tabular}
\caption{ The preferred ranges of \omeg and \ho. \label{confid} }
\end{table}

\section{Discussion}
\label{dis}

Using the Least Action Principle and a catalogue of galaxies drawn from the 
IRAS PSC-z survey, we have predicted reliable velocities for nearby galaxies.
This is consistent with the idea that PSC-z galaxies are good tracers 
of the mass at small as well as large scales. As mentioned in Section 
\ref{nam}, we are assuming galaxies are point masses and would not have 
expected to predict reliable velocities if galaxy haloes where very extended, 
or the universe contained a lot of loose matter. Consequently our
ability to predict reliable velocities on scales ranging from $2Mpc$
to $15Mpc$, argues against galaxies having extended haloes beyond $2-3Mpc$.
It is also difficult to reconcile large amounts of loose matter
with the good fit we observe. A possible solution to this is that the 
loose matter is distributed fairly evenly so that it produces no net 
accelerations. Another possibility is that the clustering 
(above galaxy scale) of the loose matter is tightly correlated 
with the clustering of galaxies. CDM models contain a lot of loose 
matter (see Section \ref{nam}), so the good fit we observe is a challenge for 
such models. Indeed as discussed in the introduction the two previous attempts
at testing the LAP in CDM simulations (Branchini \& Calberg 1994, 
Dunn \& Laflamme 1995) were not successful. It would clearly be desirable to 
repeat these investigations with an increased number of particles and
consequently higher resolution. It could then be established whether the LAP
should predict reliable velocities in CDM universes and allow an assessment 
of the effect on LAP predictions of galaxy merging, mass transfer and tidal 
disruption.

If we neglect bias (ie assume $b=1$) then the \omeg results 
confirm the difficulty of fitting the local flow for
values of \omeg around unity. Governato et al (1997) found no cases of a
CDM simulation with either 
$\Omega_{0}=0.3$ or $\Omega_{0}=1$, producing a ``cold flow'' 
as low as that observed. Their definition of ``cold flow'' was the rms velocity
inside $cz=500$kms$^{-1}$. 
They used a value of $60$\km from Schlegel et al (1994).
The data we use here gives an equivalent ``cold flow'' 
of $43\pm 35$\km, so the 
results are consistent. However our results have greater model independence,
especially with regards to the normalisation of the power spectrum. 
     
Direct comparisons of peculiar velocities have been made at larger scales,
with the use of Tully-Fisher distances. Willick et al (1997a) use $1.2 Jy$ 
galaxy redshift survey and the Mark III catalogue of peculiar velocities
to achieve an estimate of $ \beta=0.49 \pm 0.07$. da Costa et al (1998)
use the $1.2 Jy$ survey with SFI peculiar velocities to get 
$ \beta=0.6 \pm 0.1 $. The confidence limits we derive of 
$ 0.25 \leq \beta \leq 0.75$, are 
quite consistent with these estimates. For the PSC-z local weighting we get
a best fit \omeg of around $0.3$, which gives $\beta=0.5$. 
This agrees well with
Willick and da Costa estimates and suggests that there is little variation
of the measured \B with the scale the estimate is made on. Such a 
dependence of measured \B with scale has been invoked to explain the 
difference in \B estimates between velocity-velocity comparisons 
(Willick et al 1997a, da Costa et al 1998, and this paper) with 
density-density comparisons using the POTENT methodology 
(Sigad et al 1998, they estimate $ \beta=0.89 \pm 0.12 $). As we find no 
evidence for such a scale dependence, we suggest the source of the 
discrepancy must lie elsewhere. S95 use the Least Action Principle and
Tully-Fisher data to estimate $ \Omega_{0}=0.17 \pm 0.1$ (giving 
$ \beta=0.35 \pm 0.15 $, for $b=1$). Again this is consistent with our results,
but despite the similarities in methodology, our results are 
less comparable with this paper, which uses an optical 
catalogue for its mass tracers.

The classical way to estimate \hoc, has been through the distance ladder.
Typically this means using Cepheid distances to nearby galaxies, groups and
clusters to calibrate either type Ia supernova (SN Ia), or the 
Tully-Fisher relation. Direct estimates at the Virgo cluster are also made.
There tends to be a bifurcation of estimates, for example 
Freedman et al (1998) give a value 
$H_{0}=73\pm6$ (statistical) $\pm8$ (systematic) kms$^{-1}$Mpc$^{-1}$, 
whilst Tammann (1998) gives $H_{0}=56\pm7$\km Mpc$^{-1}$. 
The disagreements causing these different 
estimates are wide-ranging. Some of the more important are: the distance
to the Virgo cluster, the nature of the Fornax cluster, suitable SN Ia 
observations for calibration purposes, the use of the decline rate-
absolute magnitude relation for SN Ia and the magnitude of systematic effects
when using the Tully-Fisher relation. We obtain the confidence interval 
$65-75$ kms$^{-1}$Mpc$^{-1}$, which favours the higher estimate. 
Importantly the methodology
 we have used by-passes many of the disagreements, relying mainly on
the noncontroversial Cepheid distances. Of the disagreements listed above 
only the distance to Virgo applies to our estimate, and as shown in Section
\ref{cl}, it is of little importance to our estimate. 

Recently a number of \ho estimates from the HST Key Project have been 
presented. These Key Project estimates have been made with a variety of 
secondary distance indicators. These include: the Surface Brightness 
Fluctuations method ($H_{0}=69\pm4$ (statistical) $\pm5$ (systematic) kms$^{-1}$Mpc$^{-1}$, Ferraresa et al 1999), the Fundamental Plane relation $H_{0}=78\pm7$ (statistical) $\pm8$ (systematic) kms$^{-1}$Mpc$^{-1}$, Kelson et al 1999),
the Tully-Fisher relation ($H_{0}=71\pm4$ (statistical) $\pm10$ (systematic) kms$^{-1}$Mpc$^{-1}$, Sakai et al 1999) and the Type IA SN method $H_{0}=68\pm2$ (statistical) $\pm5$ (systematic) kms$^{-1}$Mpc$^{-1}$, Gibson et al 1999).
Perhaps the most interesting of these estimates is the Type Ia SN result 
(Gibson et al 1999). This method has traditionally favoured the lower 
\ho estimates (e.g. Saha et al 1997,1999) and thus this higher estimate removes
one of the major objections to a high value for \hoc. Our estimate for \ho ($65-75$ kms$^{-1}$Mpc$^{-1}$) agrees very well with these HST Key Project determinations.
Given that our estimate avoids many of the traditional disagreements in 
distance ladder determinations and the HST Key Project aims to provide a 
definitive estimate of \hoc, this amounts to a strong argument for a high \hoc.

Although our estimate
is made in the local volume, it is not a local estimate of \hoc. 
We use the PSC-z selection function to weight the galaxies, so if we
live in a local over- or under-density this will be fully taken
into account. As a result any net expansion or contraction of the 
local volume will be accounted for in the Least Action Predictions. 
Thus our estimate of \ho will in effect be made to at least the avergage
depth of the PSC-z ($cz = 8100$ \km) and will be a global not 
local estimate. We would like to 
reiterate that like all estimates of \ho made with Cepheids, this value is
vulnerable to errors from metallicity effects, or more importantly, an error
in the distance to the LMC.     

\section{Acknowledgements}

JS thanks PPARC for a research studentship. We thank Fabio Governato for
useful correspondence and Robert Mann for thorough readings of drafts of this 
paper, which led to great improvements.     
We also thank a referee for helpful comments which led to the paper being improved.

\end{document}